\documentclass[preprint, 3p, 12pt]{elsarticle}

\makeatletter 
\def\ps@pprintTitle{%
 \let\@oddhead\@empty
 \let\@evenhead\@empty
 \def\@oddfoot{\centerline{\thepage}}%
 \let\@evenfoot\@oddfoot}
\makeatother

\biboptions{authoryear,square}

\usepackage{enumitem}
\usepackage{placeins}
\usepackage{xfrac}
\usepackage{amsfonts}
\usepackage{bbm}
\usepackage{graphbox}

\usepackage[colorlinks]{hyperref}
\AtBeginDocument{
  \hypersetup{
    linkcolor=red,
    citecolor=blue,
    urlcolor=magenta
  }%
}%
\usepackage{color,soul}

\begin{document}

\begin{frontmatter}
\title{BundleSeg: A versatile, reliable and reproducible approach to white matter bundle segmentation}

\author[uqo]{Etienne St-Onge}
\author[drrs,vand]{Kurt G Schilling}
\author[mini]{Francois Rheault}

\address[uqo]{Department of Computer Science and Engineering, Université du Québec en Outaouais, \\Saint-Jérôme, Québec, Canada}
\address[drrs]{Department of Radiology and Radiological Sciences, Vanderbilt University Medical Center, \\Nashville, TN, USA}
\address[vand]{Vanderbilt University Institute of Imaging Science, Vanderbilt University, \\Nashville, TN, USA}
\address[mini]{Medical Imaging and Neuroinformatic (MINi) lab, Université de Sherbrooke, \\Sherbrooke, Québec, Canada}

\begin{abstract}
This work presents BundleSeg, a reliable, reproducible, and fast method for extracting white matter pathways. The proposed method combines an iterative registration procedure to a recently developped precise streamline search algorithm that enables efficient segmentation of streamlines without the need for tractogram clustering or simplifying assumptions. We show that BundleSeg achieves improved repeatability and reproducibility than state-of-the-art segmentation methods, with significant speed improvements. The enhanced precision and reduced variability in extracting white matter connections offer a valuable tool for neuroinformatic studies, increasing the sensitivity and specificity of tractography-based studies of white matter pathways.

\end{abstract}
\end{frontmatter}

\section{Introduction}
Accurate segmentation of tractography\index{tractography} bundles is crucial for advancing our understanding of brain structure and organization. Grouping streamlines into white matter (WM) bundles\index{white matter bundle} is a common practice in leveraging analysis from Diffusion-Weighted MRI (DW-MRI) models. This allows for targeted WM analysis and cohort comparison along known brain connections. Improving the reliability and reproducibility of bundle segmentation\index{segmentation} methods can effectively reduce variability in longitudinal and group analyses, thereby enhancing the overall statistical power.

Several methods have been developed to virtually dissect WM bundles from a full brain tractogram. One of the earliest approaches involves the use of region-of-interest (ROI) to virtually dissect tractography results from specific anatomical locations \citep{catani2007symmetries,oishi2008human,wakana2007reproducibility,wassermann2013describing,zhang2010atlas}. While these methods provide some control over the segmentation process, it is often limited by the subjectivity of manually drawn ROIs, or atlas resolution, and may not capture the complexity and geometry of WM pathways \citep{rheault2022influence,schilling2021tractography}. Another approach involves clustering algorithms that group streamlines based on similarity to an existing streamlines template, typically using metrics such as spatial proximity, shape, or fiber orientation \citep{garyfallidis2018recognition,o2007automatic,olivetti2017comparison}. However, most of these approaches are sensitive to initialization and may yield inconsistent results due to variability in the choice of parameters or seed points.

Additionally, machine learning-based techniques have been recently explored, where models are trained on labeled datasets to predict the presence of specific WM bundles \citep{wasserthal2018tractseg,zhang2020deep,berto2021classifyber}. Although these methods have shown promising results, they heavily rely on the availability of high-quality pre-segmented data and may be limited toward generalization to different populations or imaging protocols. Therefore, there is a need for an improved method that can address these limitations and provide a more reliable and reproducible tractography bundle.

\vspace{10pt}
To summarize, current techniques are challenged by limited repeatability (i.e. inconsistent results when run twice on the same dataset), reproducibility\index{reproducibility} (i.e. differences in scan-rescan or longitudinal acquisitions), typically due to time constraints in manual bundle segmentation or in automated segmentation methods. However, repeatable, reproducible, and anatomically accurate approaches are essential in both research and clinical tractography applications. Here, we propose a method that addresses these limitations; improving on several aspects of the tractography segmentation process in order to reduce the overall variability in bundle segmentations. This method results in more repeatable, reproducible, and anatomically accurate identification of white matter pathways, which can significantly enhance the sensitivity and specificity of tractography studies.

\section{Methods} 
In this section, we detail our proposed tractography bundle segmentation, named BundleSeg. BundleSeg is a combined procedure using an iterative registration\index{registration} technique (\ref{sec:m1}) that leverages an existing bundle distance measure (\ref{sec:m2}) and a newly proposed streamline search algorithm (\ref{sec:m3}). This integrated approach allows an efficient alignment of tractograms and segmentation of streamlines that presents strong geometric similarity to well-established brain fascicles. By incorporating these elements, our method aims to enhance both accuracy and reproducibility in tractography bundle segmentation.

\vspace{10pt}
While our template registration and segmentation strategy share similarities with existing methods such as RecoBundles\index{RecoBundles} \citep{garyfallidis2018recognition} and RecoBundlesX \citep{rheault2020analyse}, they differ at a critical step aiming to discard streamlines that are too dissimilar to the atlas. In the standard RecoBundles framework, this pruning step is performed exclusively on clusters of streamlines, which only approximates a complete streamline-to-streamline comparison. On the contrary, the pruning operation in BundleSeg is performed using an exact distance search over all streamlines. This streamline search method is described in the subsection~\ref{sec:m3}.

\subsection{Global and local registration procedure} \label{sec:m1}
The first step is to obtain a coarse registration between the native diffusion space of a subject and an atlas of bundles (generated in MNI-152). This step is performed using ANTs linear registration on anatomical images \citep{avants2008symmetric}. An exact alignment is not necessary since the next step was designed to more closely align WM pathways. After this first coarse full brain registration, each white matter bundle of the atlas is further aligned with an iterative procedure alternating between searching for the closest and most similar streamlines in the tractogram and a streamline registration algorithm. This procedure allows to gradually improve the alignment and find more accurately trajectories that are similar to the desired bundle, comparable to the streamline-based linear registration (SLR) \citep{garyfallidis2015robust}. 

\subsection{Streamlines distance} \label{sec:m2}
The minimum average direct-flip (MDF) distance is a reliable way to compute the distance between streamlines. This distance is employed to estimate how similar a streamline is to the template atlas. It has been used in various algorithms related to tractography, such as clustering, classification and outlier detection
\citep{garyfallidis2012quickbundles,garyfallidis2015robust,olivetti2017comparison,visser2011partition}. When two streamlines ($U$,$W$) have the same orientation, the MDF is equivalent to averaging the Euclidean distance along all `$m$' points ($\textbf{u}_i, \textbf{w}_i$) of those two curves: $\text{dist}(U,W) = \sum_{i=1}^m || \textbf{u}_i - \textbf{w}_i ||_2$.

\subsection{Streamlines search} \label{sec:m3}
Typically, the MDF is used to find the distance between reference streamlines (e.g. a white matter atlas) and another set of streamlines (e.g. a whole brain tractogram). However, when millions of streamlines are involved in the context of an exhaustive comparison (to find the nearest neighbor), even the most efficient distance computation will result in astronomically high computation time. Computing all possible pairs of distances would result in the MDF being computed trillions of times $\sim O(n^2)$. 

To avoid this, a K-D tree was adapted to search for nearby tractography streamlines. The resulting space partitioning tree drastically reduces the amount of computation required to find similar streamlines within a specific radius $\sim O(n \cdot log(n))$. This was possible by exploiting the Fast Streamline Search (FSS) mathematical framework recently developed to compute distances only within a maximum distance in the space of streamline \citep{st2022fast}.

As such, BundleSeg can accurately compute the bundle distance, from the WM template to all streamlines in the tractogram, at every step of the iterative procedure. This allows to avoid an approximate intermediate clustering step (in RecoBundles), or the need to better estimate the pruning distance using multiple execution (strategy from RecoBundlesX) to achieve an increased reliability. Resulting in an reproducible streamline segmentation method, with a single parameter (the maximum radius), that can exploit any existing tractography atlas.

\vspace{10pt}
The proposed segmentation procedure can be summarized by these four steps: 1) registering the subject anatomical image to the WM template reference image, 2) searching for all streamlines that resemble the bundles of interest using the resulting search tree based on the FSS framework, 3) refining the alignment of each bundle independently through SLR, and 4) iteratively repeating steps 2 and 3 while decreasing search radius until reaching the desired distance threshold.

\vspace{10pt}
The resulting search tree along with distance computation functions are available in \href{https://github.com/dipy/dipy/blob/master/dipy/segment/fss.py}{Dipy}. 
In addition, a complete segmentation pipeline improved with an exact pruning step is also available at \href{https://github.com/scilus/rbx_flow}{this repository}.

\section{Experiments}
We designed experiments in order to quantify the repeatability and reproducibility of the proposed white matter bundle segmentation algorithm, from full brain tractograms. The repeatability was assessed using a run-rerun of each algorithm on the same set of streamlines directly. The reproducibility was assessed using a scan-rescan dataset and computing the entire process end-to-end from different scans of the same subject.

\subsection{Dataset and template}
For the evaluation, we employed 43 subjects at two timepoints (scan-rescan) from the Human Connectome Project (HCP)\index{Human Connectome Project} \citep{van2013wu}. Full brain probabilistic tractograms were reconstructed using both classical local tractography and particle filtering tractography \citep{girard2014towards}, implemented in Dipy \citep{garyfallidis2014dipy}. Streamlines were generated using a WM seeding approach following fiber orientation distribution function (fODF) \citep{descoteaux2008deterministic,tournier2007robust}. Preprocessing steps included DW-MRI denoising (MRtrix) \citep{tournier2019mrtrix3}, brain extraction and tissues classification (FSL-BET, FSL-FAST) \citep{zhang2001segmentation}. It is important to note that the HCP dataset has T1-weighted images already aligned to the subject's DW-MRI space. To facilitate the analysis, all subjects' resulting streamlines were aligned to the MNI-152 space (ICBM 2009c nonlinear symmetrical) T1-weighted average \citep{fonov2011unbiased} using ANTs linear registration (antsRegistrationSyNQuick.sh) \citep{avants2008symmetric}. 

The WM bundle atlas selected for this segmentation comparison encompasses 48 bundles aligned in MNI-152 space (ICBM 2009c nonlinear symmetric) from a population average based on HCP \& UKBioBank. This template was obtained through \href{https://zenodo.org/record/7950602}{Zenodo}, developed along RecoBundlesX \citep{rheault2020analyse}. A subset of well-known bundles were used for this evaluation: Arcuate Fasciculus (AF), Corpus Callosum frontal (CC\_Fr\_2) and central portion (CC\_Pr\_Po), Cingulum (CG), Inferior Fronto-Occipital Fasciculus (IFOF), Inferior Longitudinal Fasciculus (ILF), Pyramidal Tract (PYT), Superior Longitudinal Fasciculus (SLF).

\subsection{Evaluation}
To assess the repeatability and reproducibility of the proposed tractography bundle segmentation method, a total of two full brain tractograms were computed for each 43 HCP scan-rescan subjects, one at each session. This was done to determine the scan-rescan reproducibility for each compared method. In addition, the run-rerun variability was estimated by executing each bundle segmentation algorithm twice using identical streamlines and parameters, but with a distinct random number generator.

Multiple measures were used to evaluate each bundle segmentation approach: the bundle volume, the number of streamlines, and the average streamline length. To estimate the variability, these measures were compared in both run-rerun and scan-rescan segmentations using an absolute difference (L1-norm) averaged over all subjects, along with the standard deviation (±).

Volumetric Dice coefficient\index{Dice coefficient} was computed for both run-rerun and scan-rescan, to evaluate the overall volume similarity, as well as voxel overlap and overreach. For non-overlapping voxels, the average distance to the nearest corresponding voxel was computed, describing the ``adjacency'' distance between two segmentations. Compared to the overreach percentage in the Dice coefficient, this describes how distant on average are two successive bundle extractions.

BundleSeg computation time was compared to the standard RecoBundles as well as a multi-parameters RecoBundles with atlas fusion (RecoBundlesX) algorithms, using an Intel Skylake 6148 at 2.4 GHz.

\section{Results}
\subsection{Computation time}
Computation times per subject for each method were: 18.52 ± 3.09 minutes (RB), 124.85 ± 21.60 minutes (RBX), 8.10 ± 1.52 minutes (BundleSeg, proposed). BundleSeg achieves a remarkable increase in reproducibility and overall quality compared to RecoBundles (RB) for an execution twice (2.3x) as fast. BundleSeg surpasses RecoBundlesX (RBX) in reproducibility and overall quality with an execution approximately 15 times faster. Our proposed method is faster than both baselines while offering more reproducible and better quality segmentations in both run-rerun and scan-rescan settings, qualitative and quantitative analysis of segmentation results are presented below. The computation time includes loading, clustering, streamlines search and distance computation, and finally saving. This loading, clustering and saving are the bulk ($\sim$70\%) of the computation time for BundleSeg.

\begin{figure}[!t]
  \centering
  \includegraphics[trim=25 0 25 0, clip, width=0.99\linewidth]{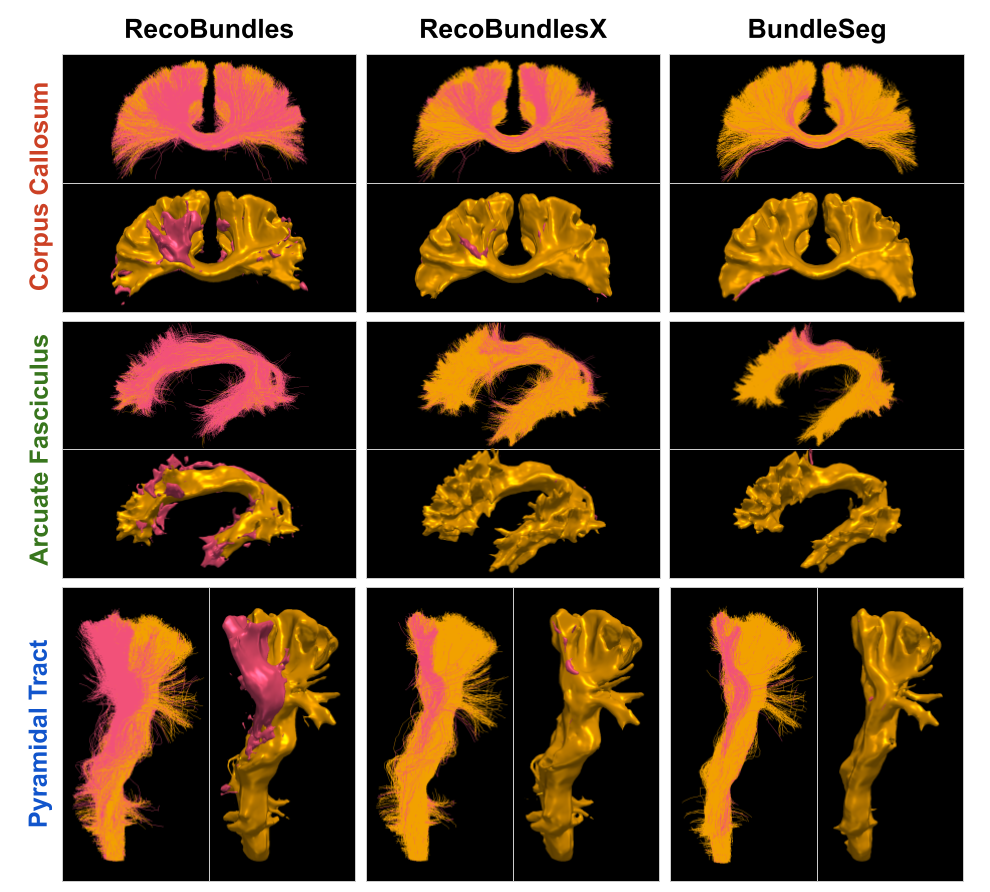}
  \caption{Run-rerun visual comparison of two successive segmentations with the exact same streamlines (different random seeds), between two baselines and the proposed method over 3 bundles. Bundle streamlines and voxels overlap are displayed in orange (consistent in both run-rerun segmentations) and overreach in pink (inconsistent).}
  \label{fig:i1}
\end{figure}

\subsection{Qualitative comparison, segmentation overlap \& overreach}
Figure~\ref{fig:i1} visualizes streamlines and volumetric results for the run-rerun analysis for RecoBundles (RB), RecoBundlesX (RBX), and the proposed method (BundleSeg) from a single randomly chosen subject. Areas of overlap (in orange) and difference (in pink) are shown for each bundle and each algorithm for both streamlines and voxel-wise. While all algorithms (for all pathways) show similar locations, shape, and size of pathways when rerun on the same data, RB often recognizes and segments very different streamlines when run twice on the same algorithm, which can result in different estimated pathway volumes. This run-rerun variation is considerably reduced by RBX, and nearly non-existent with BundleSeg.

\begin{figure}[!b]
  \centering
  \includegraphics[trim=0 15 0 15, clip, width=0.92\linewidth]{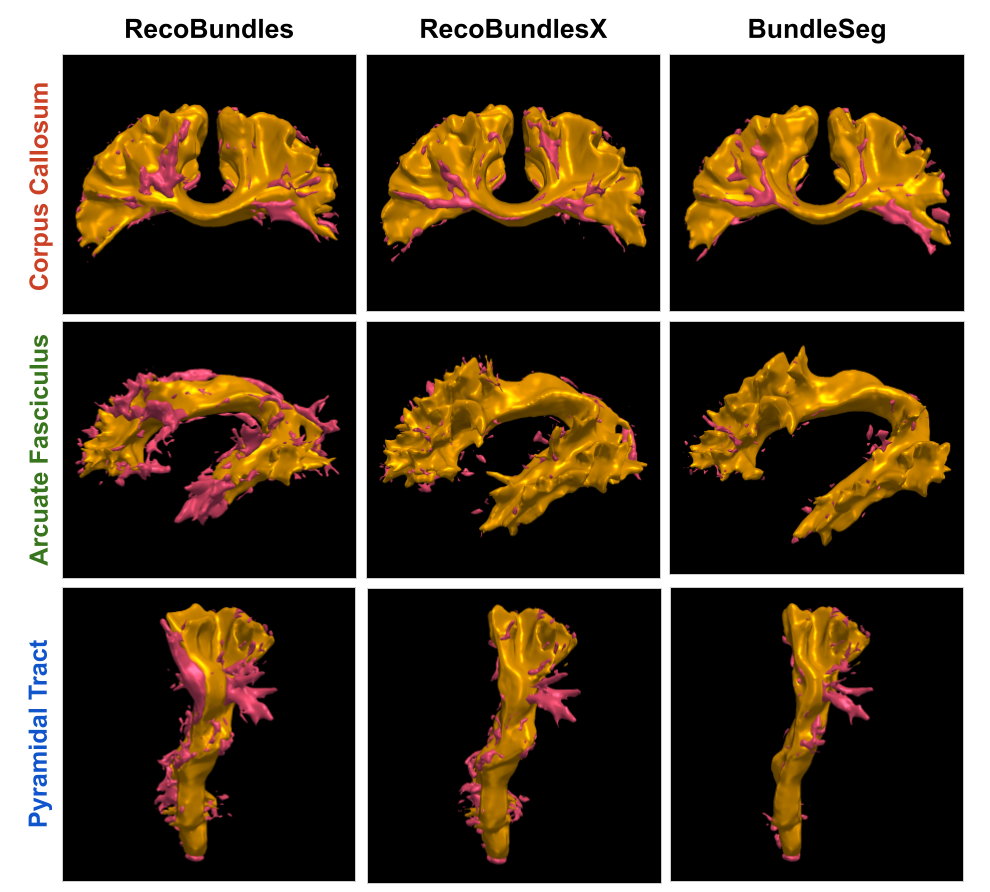}
  \caption{Scan-rescan visual comparison of two independently reconstructed tractograms (from distinct acquisitions), between two baselines and the proposed method on three bundles. Bundle volume overlap is displayed in orange (consistent in both scan-rescan segmentations) and overreach in pink (inconsistent).}
  \label{fig:i2}
\end{figure}

Figure~\ref{fig:i2} shows the volumetric overlap-overreach in a scan-rescan setting, with two independent segmentations from different DW-MRI acquisitions. Again, all algorithms result in visually similar scan-rescan segmentations, but there are variations, particularly at the edge of bundles, with more variation in RB, followed by RBX, and BundleSeg.

\subsection{Quantitative results, reproducibility and variability}
\begin{figure}[!t]
  \centering
  \includegraphics[trim=10 0 5 0, clip, width=0.98\linewidth]{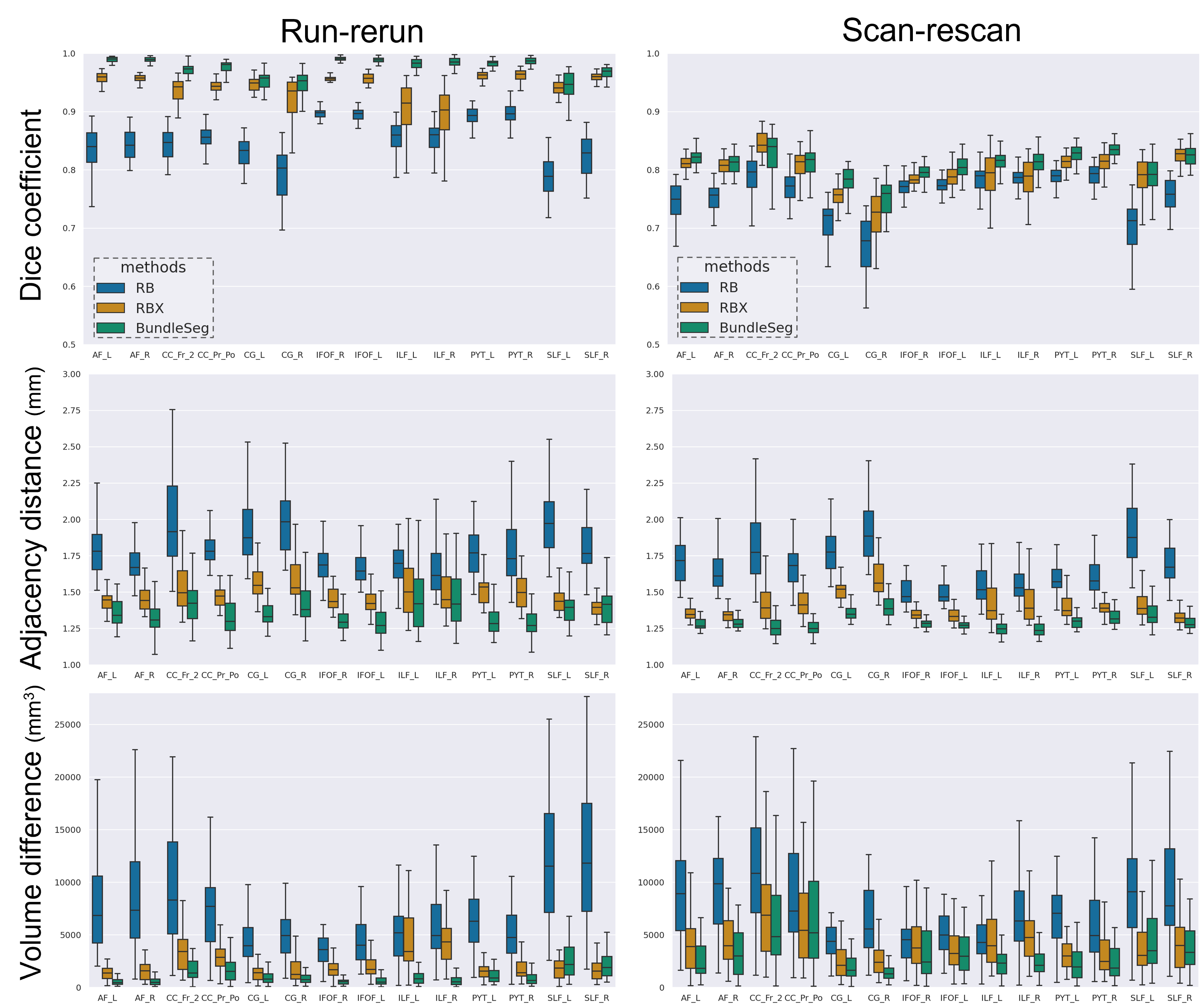}
  \caption{Comparison of agreement in both run-rerun (left) and scan-rescan (right) for 3 measures: Dice coefficient (top row, score of 1 is a perfect match) Adjacency, millimetric distance for non-overlapping voxel (middle row, lower is better with a minimum of 1 $mm$) and Difference in volume in $mm^3$ (bottom row, 0 is better). We compared the standard RecoBundles framework, the multi-parameters with atlas fusion RecoBundles framework (RecoBundlesX) and our proposed method. A sample of representative major WM pathways were selected to showcase the agreement measures. The proposed BundleSeg surpasses RecoBundles and RecoBundlesX framework is nearly all bundles, for all 3 agreement measures for both the run-rerun and scan-rescan. The average scores are improved, and the standard deviation is decreased.}
  \label{fig:graphs}
\end{figure}

Figure~\ref{fig:graphs} shows quantitative run-rerun and scan-rescan results, for all three methods, where several trends are apparent. First, for almost every pathway the Dice coefficient shows that BundleSeg outperforms both the standard RecoBundles framework, as well as RecoBundlesX for most bundles (while equal for others), in both run-rerun and scan-rescan. In run-rerun, BundleSeg results in a near-perfect volumetric match ($>$ 0.95) in all tested pathways. Scan-rescan shows an expected decreased volume overlap for all methods when compared to run-rerun results. Next, BundleSeg has the lowest adjacency distances in both run-rerun and scan-rescan, and RBX is second. Importantly, adjacency distance is around one or two voxels ($mm$) on average, for all algorithms. Finally, differences in volume can largely vary when using RB, even on the same tractogram. Repeatability and reproducibility of volume can be strongly improved through RBX, and further enhanced with BundleSeg.

\section{Discussion}
In this work, we introduced a bundle segmentation algorithm that overcomes challenges associated with existing manual and automated machine learning-based segmentation techniques. This method, named BundleSeg, resulted in a significant computational speedup, along with improved repeatability and reproducibility compared to current state-of-the-art methods. 

While RecoBundlesX has much better reproducibility than its predecessor RecoBundles, it requires multiple comparisons, resulting in higher computation time. On the other hand, the proposed method is 2x and 15x faster than RecoBundles and RecoBundlesX respectively. This can be attributed to the fast streamlines search making the distance computation and pruning an order of magnitude faster than the clustering approach of RecoBundles and RecoBundlesX.

The proposed method displayed an improvement in both scan-rescan reproducibility compared to existing approaches. This can be observed qualitatively in Figures~\ref{fig:i1}-\ref{fig:i2}, and quantitatively in all graphs from Figure~\ref{fig:graphs}. BundleSeg has a significantly higher Dice coefficient, indicating a greater agreement between the repeated segmentations, more specifically in run-rerun where it has a near-perfect Dice along with a very good adjacency. For run-rerun comparison, a ``Streamline Dice'' coefficient was also computed describing the overlap of streamlines (as the element of comparison instead of voxels); but was not included since results were similar to ``Voxel Dice'' results.

Furthermore, our results exhibited a notable reduction in volume variability. This lower same-subject absolute volume difference in WM pathway segmentations indicates that our method consistently extracted a similarly shaped bundle. The reduced variability indicates the stability and reliability of our approach, minimizing the influence of random initialization or other factors that could introduce variability.

All our results suggest that the proposed algorithm reduces the inherent variability associated with tractography bundle segmentation, enabling more consistent and reproducible results across multiple sessions. While the reduced Dice alignment in scan-rescan is lower, part of it might be caused by the tractography reconstruction variability.

\vspace{10pt}
Future work should include more comparison to other segmentation methods, based on traditional classification and deep learning. Second, it remains to be seen whether the higher precision and improved reproducibility lead to more sensitive detection of white matter changes in disease and disorder. And finally, it is worth exploring how this exhaustive and fast streamline search\index{fast streamline search} can improve longitudinal analysis on subjects with multiple timepoints.

\section{Conclusion}
In this work, we proposed a reliable and robust approach for extracting white matter pathways. The novelty of BundleSeg resides in a precise search algorithm, which efficiently identifies all relevant tractography streamlines corresponding to specific white matter bundles. Using an exact and exhaustive streamline radius search, instead of an approximation, ensures that the segmentation process is comparable to previous work while significantly improving speed and reliability. Furthermore, combining it with an iterative registration results in an overall higher reproducibility and lower variability. This improved stability provided by our method has important implications for neuroimaging studies, enabling researchers to obtain more reliable and robust results when investigating white matter connectivity patterns and their associations with various clinical or cognitive variables.

\section*{Conflict of Interest} 
We have no conflict of interest to declare.

\FloatBarrier
\bibliography{bibfile}
\end{document}